\documentclass[useAMS,usenatbib]{mn2e}
\usepackage[utf8]{inputenc}
\usepackage{journal_names}
\usepackage{graphicx}
\usepackage{natbib}
\usepackage{times}
\usepackage{url}
\usepackage{amssymb}
\usepackage{color}

\newcommand{\Pp}{P_{1.4\mbox{\,GHz}}}
\newcommand{\WHz}{W\,Hz$^{-1}$}

\title[Radio-AGN Feedback]{Radio-AGN Feedback: When the Little Ones were Monsters}
\author[Williams et al.]{W.~L.~Williams\thanks{E-mail: wwilliams@strw.leidenuniv.nl (WLW)}$^{1,2}$ and H.~J.~A.~R\"{o}ttgering$^{1}$\\
$^{1}$Leiden Observatory, Leiden University, P.O. Box 9513, 2300 RA Leiden, The Netherlands\\
$^{2}$Netherlands Institute for Radio Astronomy (ASTRON), P.O. Box 2, 7990AA Dwingeloo, The Netherlands
}
\begin{document}

\maketitle
\begin{abstract}
We present a study of the evolution of the fraction of radio-loud active galactic nuclei (AGN) as a function of their host stellar mass. We make use of two samples of radio galaxies: one in the local universe, $ 0.01 < z \leq 0.3$, using a combined SDSS-NVSS sample and one at higher redshifts, $0.5 < z \leq 2$, constructed from the VLA-COSMOS\_DEEP Radio Survey at $1.4$~GHz and a K$_{s}$-selected catalogue of the COSMOS/UltraVISTA field. We observe an increase of more than an order of magnitude in the fraction of lower mass galaxies ($M_* < 10^{10.75}$\, M$_{\sun}$) which host Radio-Loud AGN with radio powers $\Pp > 10^{24}$\,\WHz at $z \sim 1-2$ while the radio-loud fraction for higher mass galaxies  ($M_* > 10^{11.25}$\,M$_{\sun}$) remains the same. We argue that this increase is driven largely by the increase in cold or radiative mode accretion with increasing cold gas supply at earlier epochs. The increasing population of low mass Radio-Loud AGN can thus explain the upturn in the Radio Luminosity Function (RLF) at high redshift which is important for understanding the impact of AGN feedback in galaxy evolution.
\end{abstract}


\section{Introduction}
\label{sect:intro}

During recent years it has become increasingly apparent that the radio jets of radio-loud Active Galactic Nuclei (RL AGN or radio-AGN) play a crucial role in the process of galaxy formation and evolution via `AGN feedback' \citep[see e.g.][]{Best.et.al.2006, Best.et.al.2007,Bower.et.al.2006,Croton.et.al.2006,Fabian.et.al.2006}. This feedback can occur because the enormous energy output of the AGN can be injected into the surrounding medium, possibly also the fuel source of the AGN, via ionizing radiation and/or relativistic jets, thereby providing enough energy to affect star formation in the host galaxy. {Radio galaxies have been shown to comprise two different populations: high and low excitation \citep{Best.et.al.2005a, Tasse.et.al.2008, Hickox.et.al.2009}, each of which may have a separate and different effect of feedback, the exact nature and evolution of which is still debated. The first population of RL AGN is associated with the classic optical `quasars'. These sources radiate across the electromagentic spectrum and are consistent with the unified models of quasars where emission is obscured at some wavelengths when the source is viewed edge-on  \citep[e.g.][]{Barthel.1989,Antonucci.1993,Urry.Padovani.1995}. In this `high-excitation', `cold mode'  or `radiative mode', accretion is postulated to occur via an accretion disc in a radiatively efficient manner \citep[e.g.][]{Shakura.Sunyaev.1973}. These high excitation radio galaxies (HERGs) are typically hosted by lower mass, bluer galaxies in less dense environments \citep[e.g.][]{Tasse.et.al.2008,Janssen.et.al.2012}. The second mode of radio activity was first noted by their lack of strong high-excitation narrow-line optical excitation expected from the `quasar' mode \citep{Hine.Longair.1979,Laing.et.al.1994,Jackson.Rawlings.1997}. Moreover they show no evidence of mid-infrared emission from dusty tori \citep{Whysong.Antonucci.2004,Ogle.et.al.2006} and no evidence of accretion-related X-ray emission \citep{Hardcastle.et.al.2006,Evans.et.al.2006}.  \cite{Hardcastle.et.al.2007} first suggested that this mode, known as the `low-excitation', `radio mode', `hot mode' or `jet mode' occurs when hot gas is accreted directly onto the supermassive black hole in a radiatively inefficient manner via advection dominated accretion flows \citep[ADAFs, e.g.][]{Narayan.Yi.1995}.   \cite{Best.et.al.2005a} showed that these low excitation radio galaxies (LERGs) are hosted by fundamentally different galaxies: higher mass, redder and occurring in more dense environments. This mode in particular provides a direct feedback connection between the AGN and its hot gas fuel supply in the manner of work done by the expanding radio lobes on the hot intra-cluster gas. For a more detailed review of the HERG versus LERG dichotomy see \cite{Heckman.Best.2014} and references therein. }

In order to understand the relative significance of the different types of radio-AGN feedback we need to understand the cosmic evolution of radio sources in detail. It has been known for several decades that the comoving number density of powerful radio sources is  two to three orders of magnitude greater at a redshift of two to three compared to the local Universe \citep[e.g.][]{Schmidt.1968,Sandage.1972,Osmer.1982,Peacock.1985,Schmidt.Schneider.Gunn.1988,Dunlop.Peacock.1990,Rigby.et.al.2011}. Similarly, the space density of optically selected quasars (QSOs) peaks at redshift $2< z <3$ \citep[e.g.][]{Boyle.1988,Hewett.1993,Warren.et.al.1994}.  It is also well known that, within the local universe ($z \la 0.3$), the  fraction of galaxies which host a radio source, i.e. the radio-loud fraction, is a very steep function of host galaxy stellar mass \citep[$f_{\mbox{radio-loud}}\propto M_*^{2.5}$,][]{Best.et.al.2005b}, increasing to  $>30$~per~cent at stellar masses above $5\times10^{11}$\,M$_{\sun}$ for radio luminosities $>10^{23}$\,\WHz. This pervasiveness of radio-loudness among high mass galaxies in the local Universe, combined with the dramatic increase in density of radio-loud sources at earlier times suggests that there must be an increase in the prevalence of radio activity among galaxies of lower mass.  To test this, we investigate the fraction of radio-loud sources out to redshift $\sim2$ as a function of their host stellar mass.

This paper is structured as follows. In Section~\ref{sect:samples} we describe the construction of the Radio-Loud AGN samples and the catalogues from which they are selected. In Section~\ref{sect:LF} the luminosity functions for these samples are constructed and binned by host galaxy stellar mass. The radio-loud fraction as a function of host stellar mass is also determined. We discuss the results and their implications in Section~\ref{sect:interpret} and conclude in Section~\ref{sect:conclude}. Throughout this paper, we use the following cosmological parameters: $H_0 = 70$\,km\,s$^{-1}$\,Mpc$^{-1}$, $\Omega_m = 0.3$ and $\Omega_\Lambda = 0.7$. The spectral index, $\alpha$, is defined as $S_{\nu} \propto \nu^{-\alpha}$ and unless otherwise specified, we adopt a default value of $0.8$.

\begin{figure*}
 \centering
 \includegraphics[width=0.45\textwidth]{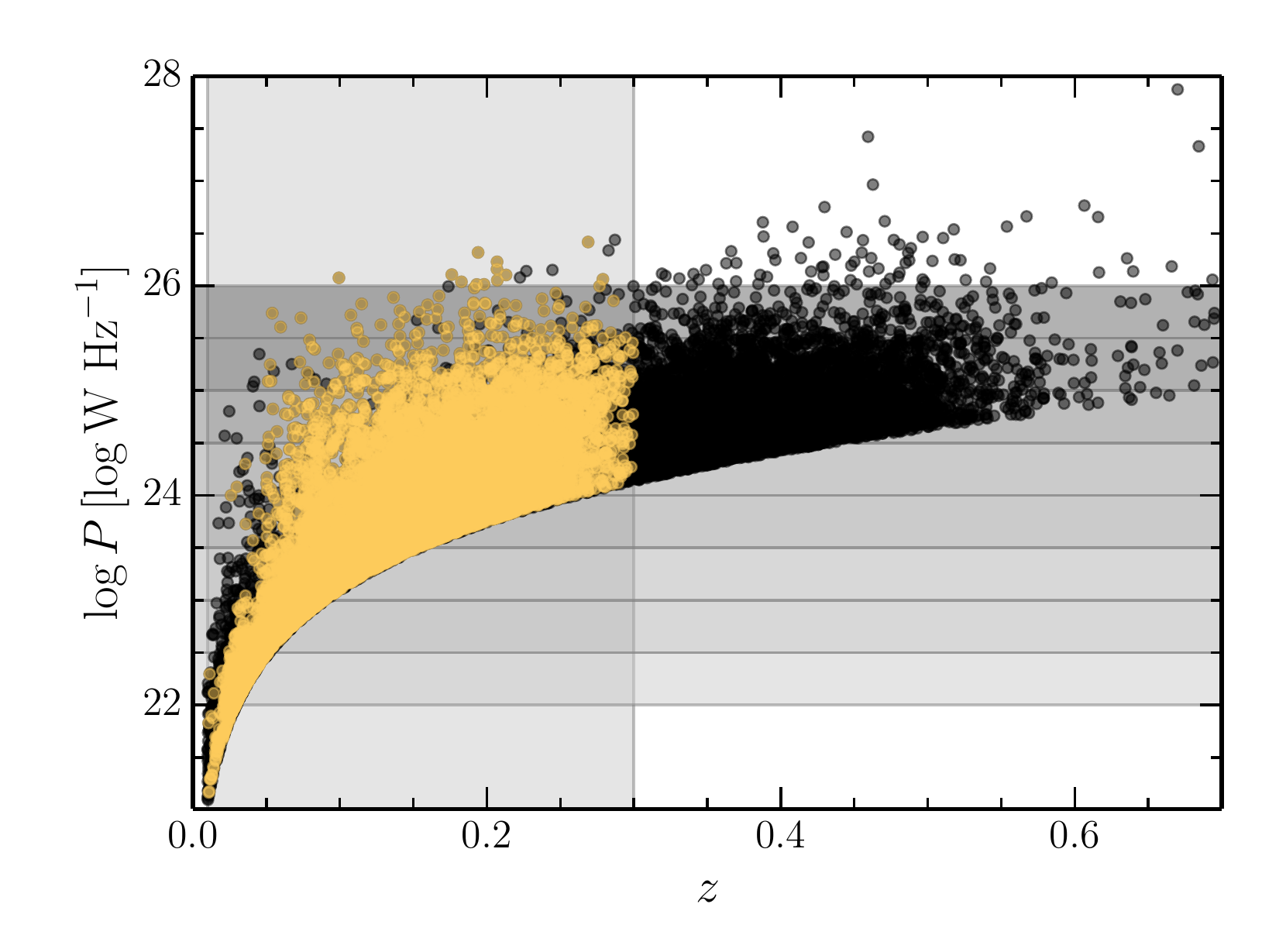}
 \includegraphics[width=0.45\textwidth]{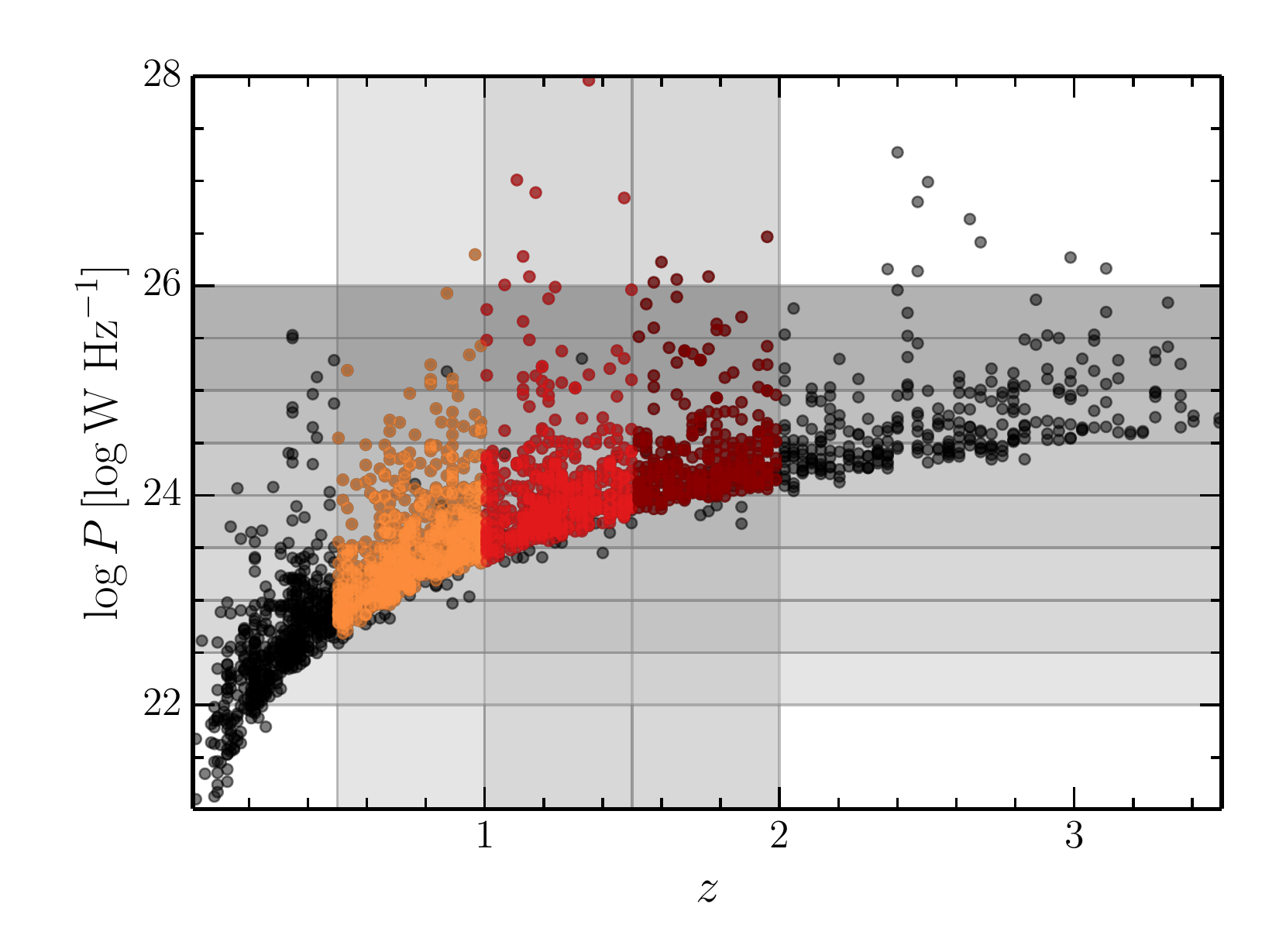} \\
 \includegraphics[width=0.45\textwidth]{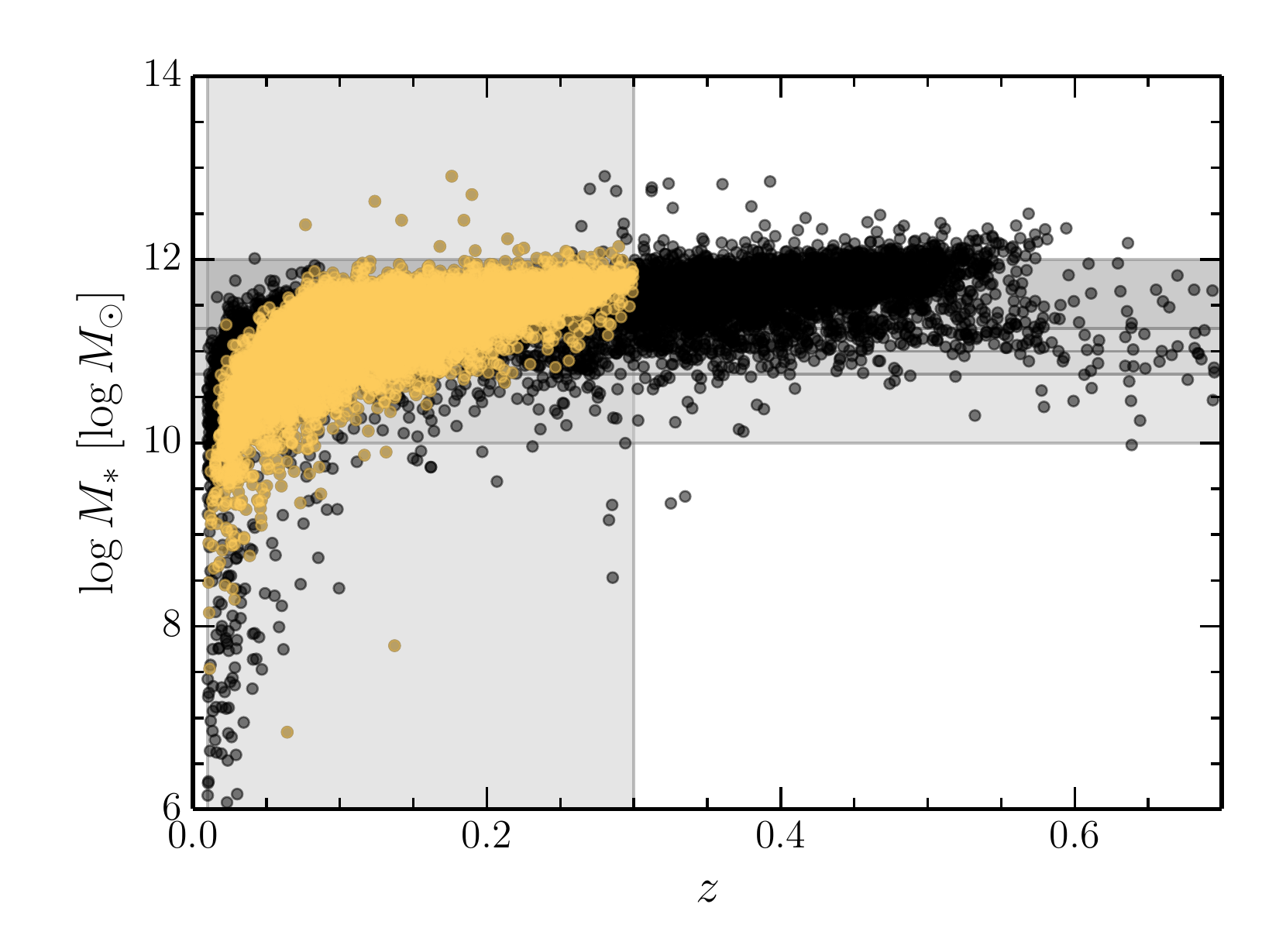}
 \includegraphics[width=0.45\textwidth]{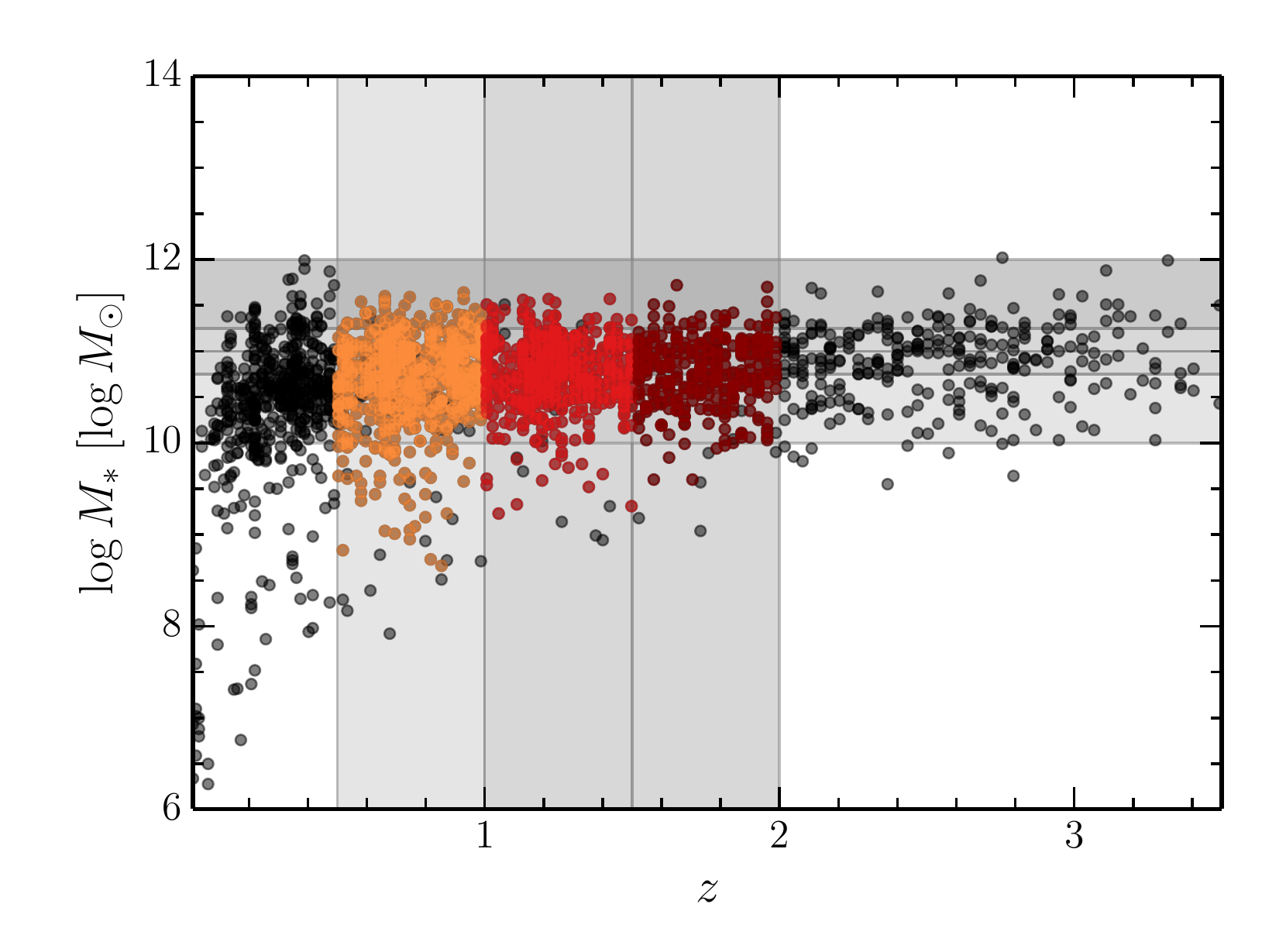} \\
 \caption{(left) Full SDSS-NVSS Sample and the selected `local' $0.01 <z \leq 0.3$ sample : radio power vs $z$ and stellar mass, $M_*$, vs z. (right) Full VLA-COSMOS Sample and the three selected `high redshift' $0.5 <z \leq 1$ and $1 <z \leq 2$ samples : radio power vs $z$ and stellar mass, $M_*$, vs $z$.  The grey shaded regions show the regions in mass-space, redshift-space and radio power-space which demarcate the redshift samples, mass bins and power bins for the Luminosity Functions. The selected clean mass-complete radio samples are shown in yellow, orange and red symbols (these colours denote these redshift bins thought this paper).}
 \label{fig:samplesboth}
\end{figure*}

\section{Radio-Loud AGN Samples}
\label{sect:samples}

To investigate the evolution of the Radio Luminosity Function (RLF) for different stellar-masses over a similar luminosity range several samples with good ancillary and derived data are needed, to provide both the required statistics at low redshift and the sensitivity at high redshift. In this work we combine one already existing matched radio-optical dataset in the local universe using the SDSS-NVSS sample, described in Section~\ref{sect:sdsssample}, and one at redshifts $0.5 < z < 2$, which we have constructed using the VLA-COSMOS DEEP Radio Survey at $1.4$~GHz and a  K$_{s}$-selected catalogue of the COSMOS/UltraVISTA field, which is described in Section~\ref{sect:cosmossample}.

\subsection{Local SDSS Sample}
\label{sect:sdsssample}

For the local radio source sample we use the catalogue compiled by \cite{Best.Heckman.2012}, which was  constructed by cross-matching optical galaxies from the seventh data release \citep[DR7;][]{Abazajian.et.al.2009} of the Sloan Digital Sky Survey (SDSS) spectroscopic sample with radio sources in  the NRAO Very Large Array (VLA) Sky Survey \citep[NVSS;][]{Condon.et.al.1998} and the Faint Images of the Radio Sky at Twenty centimetres \citep[FIRST;][]{Becker.et.al.1995}. The parent optical sample consists of all galaxies in the value-added spectroscopic catalogues (VASC) created by the Max Plank Institute for Astrophysics and Johns Hopkins University (MPA-JHU) group \citep[see][available at \url{http://www.mpa-garching.mpg.de/SDSS/}]{Brinchmann.et.al.2004}. The cross-matching was done for all radio sources with flux densities $>5$\,mJy, which corresponds to radio luminosities of $\Pp  \gtrsim 10^{23}$\,\WHz at redshift $z=0.1$ and $\Pp \gtrsim 10^{24}$\,\WHz at redshift $z=0.3$. The combined radio-optical area covered is $2.17$\,str \citep{Best.Heckman.2012}. Of the $927\,522$ galaxies in the VASC, \cite{Best.Heckman.2012} selected a magnitude-limited sample of $18\,286$ radio sources, which they showed to be $95$~per~cent complete and $99$~per~cent reliable \citep{Best.et.al.2005a}. The sample was restricted to the `main galaxy sample' \citep{Strauss.et.al.2002}, comprising galaxies within the magnitude range $ 14.5 \leq r < 17.7$\,mag and the redshift range $ 0.01 < z \leq 0.3$. This local radio-optical sample consists of $9168$ radio sources. We note that, being based on the SDSS main galaxy sample, this local matched radio-optical sample excludes both radio-loud quasars and broad-line radio galaxies. However, \cite{Best.et.al.2014} show that this {is} only problematic at radio powers above $\Pp \gtrsim 10^{26}$\,\WHz. Since our LFs do not probe those high powers we make no correction for this bias. Moreover, we know that the dominant population of radio sources in this sample are not quasars.

Properties of the host galaxies are taken from the VASCs which, for each source, provide several basic measured parameters from the imaging data such as magnitudes, colours and sizes \citep{York.et.al.2000}, as well as derived properties including, most importantly for this work, the stellar mass $M_*$ \citep{Kauffmann.et.al.2003b}. For their matched radio sample, \cite{Best.Heckman.2012} also separated the sources into star-forming galaxies and RL AGN ($7302$ sources), which are further sub-divided into high-excitation (HERG) and low-excitation (LERG) sources, based on their optical photometric and spectroscopic parameters.

The left panels in Fig.~\ref{fig:samplesboth} show the radio power and host stellar mass as a function of redshift for the radio sources in the SDSS-NVSS sample. All the radio sources are plotted in black, and sources in the restricted sub-sample within the redshift bin which defines our  local sample are plotted in yellow.

\subsection{Distant VLA-COSMOS Sample}
\label{sect:cosmossample}

The VLA-COSMOS Large Project covered the $2$\,deg$^2$ of the COSMOS Field at $1.4$\,GHz with observations by the VLA in the A configuration. This survey, extensively described in \cite{Schinnerer.et.al.2004, Schinnerer.et.al.2007}, provides continuum radio observations with a resolution (half-power beam width) of $1.5 \times 1.4$\,arcsec and a mean $1\sigma$ sensitivity of about $10.5$\,$\mu$Jy in the innermost $1$\,deg$^2$ region and of about $15$\,$\mu$Jy in the outer parts. From the VLA-COSMOS catalogue we selected sources above $50\,\mu$Jy, which corresponds to sources  $>5\sigma$ over $50$~per~cent of the survey area.

The optical data for this sample comes from the K$_{s}$-selected catalogue of the COSMOS/UltraVISTA field \citep{Muzzin.et.al.2013} which contains PSF-matched photometry in $30$ photometric bands covering the wavelength range $0.15\micron \rightarrow 24\micron$. The entire region overlaps with the radio survey so the combined area is that of the COSMOS/UltraVISTA data, $1.62$\,deg$^2$.  Following the recommended criteria of \cite{Muzzin.et.al.2013}, we selected a `clean' sample from the $K_s$ catalogue of sources with flags: `star' $\neq 1$, `K\_flag' $\leq 4$, `contamination' $ \neq 1$, and `nan\_contam'$\leq 3$ (these flags relate to stars and saturated sources and the quality of the photometry for nearby sources). Finally we selected sources brighter than the $90$\ per\ cent completeness limit of $K_s \leq 23.4$\,mag. The redshifts provided are determined from SED fitting to the broadband photometry using the EAZY code \citep{Brammer.et.al.2008} and for source $z<1.5$ they quote an rms error of $(\delta z / (1+z)) = 0.013$ and a catastrophic outlier fraction of $1.56$\,per\ cent.  While the quoted outlier fraction is very low, we do expect that many of the outliers will be quasars (i.e. HERGs). It is well known that photometric redshifts determined for quasars are less reliable than those obtained for galaxies \citep[e.g.][]{Richards.et.al.2001,Babbedge.et.al.2004,Mobasher.et.al.2004,Polletta.et.al.2007}.

Within this sample we defined three `high-redshift' sub-samples in the redshift ranges $0.5 < z \leq 1.0$, $1.0 < z \leq 1.5$ and $1.5 < z \leq 2.0$, chosen such that each contains $\sim300-700$ galaxies. The right panels of Fig.~\ref{fig:samplesboth}  show the radio power and host stellar mass as a function of redshift for the radio sources in the VLA-COSMOS sample.  Sources in the restricted ranges which define our high redshift samples are plotted in orange and red, and all the remaining sources are shown in black.

Note that at radio powers $\Pp \gtrsim 10^{23}$\,\WHz, the samples should consist almost entirely of RL AGN -- we expect very  little contamination from star-forming galaxies. We confirmed this using the spectroscopy-based AGN/SF separation in the local sample \citep{Best.et.al.2005b}. Moreover, the star formation rate for this radio power is in excess of $25$\,M$_{\sun}$\,yr$^{-1}$  \citep{Condon1992} so  we expect AGN to continue to dominate even at higher redshifts.

\section{The Stellar-Mass dependent Luminosity Function}
\label{sect:LF}

\subsection{The Luminosity Function}

A standard technique for quantifying the rate of evolution of a population of galaxies is to compare their luminosity functions (LFs) at two different epochs. In this section, we therefore determine the evolution of the Radio-Loud AGN population among host galaxies of different masses by comparing the luminosity functions of the SDSS-NVSS and VLA-COSMOS samples. Determining the radio LFs for the full samples first serves to confirm our sample selection and methods.

These radio luminosity functions were calculated in the standard way, using $\rho = \Sigma_{i} 1/V_i$ method \citep{Schmidt.1968,Condon.1989}, where $V_i$ is the volume in which a given source could be detected. This volume is determined by both the minimum and maximum distance at which a given source would be included in the sample as a result of the selection criteria:  $V_i = V_{\mbox{max}} - V_{\mbox{min}}$, where $V_{\mbox{max}}$ and $V_{\mbox{min}}$ are the volumes enclosed within the observed sky area out to the maximum and minimum distances respectively. The minimum accessible volume is a result of the  optical cut-off on the bright end  ($>14.5$\,mag for SDSS-NVSS, while the VLA-COSMOS sample has no bright flux limit). The maximum accessible volume is determined by the flux limits of both the optical ($<17.77$\,mag for SDSS-NVSS and $<23.4$\,mag for VLA-COSMOS) and radio data ($>5$\,mJy for SDSS-NVSS and $>50\,\mu$Jy for the VLA-COSMOS sample). In practice, the maximum redshift is dominated by the radio flux limits.
The normalization of the luminosity function requires knowledge of the precise intersection area of all input surveys. The sky area for the SDSS-NVSS sample is taken to be $2.17$\,sr \citep{Best.et.al.2005b}. For the  VLA-COSMOS sample, the sky area is taken to be $1.62$\,deg$^2$ for all sources above $100\,\mu$Jy. For the faintest sources in the VLA-COSMOS sample, the area in which the sources could be detected is smaller due to the non-uniform rms noise level in the VLA-COSMOS mosaic. We therefore weight each source by the inverse of the area in which it can be detected, which also accounts for the varying detection area within a given luminosity bin. {Uncertainties are calculated as the statistical Poissonian errors with a contribution of cosmic variance where appropriate; in some luminosity/mass bins, these errors are so small that the error will be dominated by systematic effects. The area covered by VLA-COSMOS sample of $1.62$\,deg$^2$ is small enough that the effects of cosmic variance are  not negligible for very
massive galaxies. The total area covered by COSMOS/UltraVISTA is approximately 1.5 square degrees in one single field, making the effects of cosmic variance not negligible for very
massive galaxies. The contribution of cosmic variance to the total error budget was estimated through the recipe of \citep{Moster.et.al.2011}. The average uncertainties due
to this effect for the least massive galaxies vary from $6$ to $7$~per~cent and for the most massive galaxies vary from $10$ to $12$~per~cent  across the three highest redshift bins. These values were
added in quadrature to the Poissonian error of the LFs.}

\begin{figure}
 \centering
\includegraphics[width=0.49\textwidth]{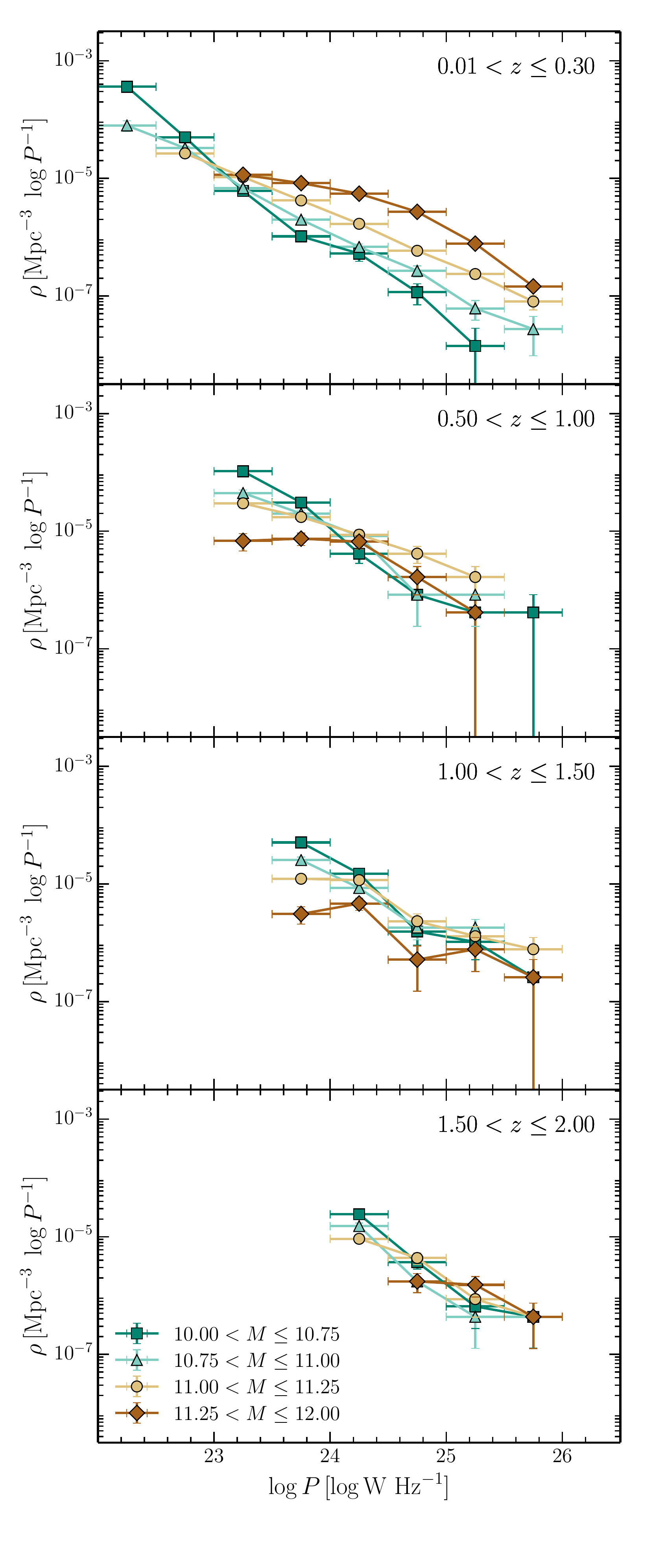} \\
 \caption{The comoving space density of radio-loud sources. This is the RLF in redshift bins $ 0.01 < z \leq 0.3$ (top panel), $0.5 < z \leq 1.0$ (top-middle panel), $1.0 < z \leq 1.5$ (bottom-middle panel) and $1.5 < z \leq 2.0$ (bottom panel) for the four host stellar mass bins (plotted in colour).}
 \label{fig:RLF_in_massbins}
\end{figure}

\begin{table}
 \begin{center}
 \caption{Number of sources in each stellar mass--redshift bin. }
 \label{tab:samplesizes}
\begin{tabular}[c]{lcccc}
\hline
\multicolumn{1}{c}{Mass bin} & \multicolumn{4}{c}{Redshift bin} \\
\multicolumn{1}{c}{$\log (M_*/M_{\odot})$} & $ 0.01 - 0.3$ & $0.5 - 1.0$ & $1.0 - 1.5$ & $1.5 - 2.0$ \\
\hline
$ 10.00 - 10.75 $ & 1021 & 296 &  231&  147\\
$ 10.75 - 11.00 $ & 843 & 153 & 126 &  90\\
$ 11.00 - 11.25 $ & 1639 & 127 & 111 &  74\\
$ 11.25 - 12.00 $ & 5306 & 51 & 35 &  26\\
Total & 9006 & 672 & 518 &  343\\
\hline
 \end{tabular}
 \end{center}
\end{table}

We have compared our derived RLFs for radio-loud sources of all stellar masses with those in the literature. For the full samples, our local RLF agrees with that of \cite{Best.et.al.2005a} and our RLF for the high redshift VLA-COSMOS sample agrees with that of \cite{Smolcic.et.al.2009}. The agreement is not unexpected as we are using the same data, but it serves to validate out sample selections and method.

We next split the samples into four bins of different host galaxy stellar masses: $ 10.0 <  \log (M_*/M_{\odot}) \leq 10.75 $, $ 10.75 <  \log (M_*/M_{\odot}) \leq 11.0 $, $ 11.0 <  \log (M_*/M_{\odot}) \leq 11.25$, and $ 11.25 <  \log (M_*/M_{\odot}) \leq 12.0 $. These bins are chosen to sample the stellar masses well with similar numbers of sources in each bin (see Fig.~\ref{fig:samplesboth}). The RLFs derived in each of these stellar mass bins in the four redshift bins are plotted in Fig.~\ref{fig:RLF_in_massbins}. Note that in the low redshift sample we exclude the points below $\Pp \lesssim 10^{23}$\,\WHz for the highest stellar mass bin as these sources are only detectable out to $z \lesssim 0.1$ and this sample is incomplete for high mass sources at these redshifts. Table~\ref{tab:samplesizes} shows the number of sources in each stellar mass-redshift bin. From these LFs we see that in the local universe and for  $\Pp \gtrsim 10^{23}$\,\WHz, the number density of the highest host stellar mass bin is the greatest, while in the higher redshift bins the number density within all the stellar mass bins is becoming increasingly similar. 

\subsection{Space Density Evolution}

\begin{figure}
 \centering
 \includegraphics[width=0.49\textwidth]{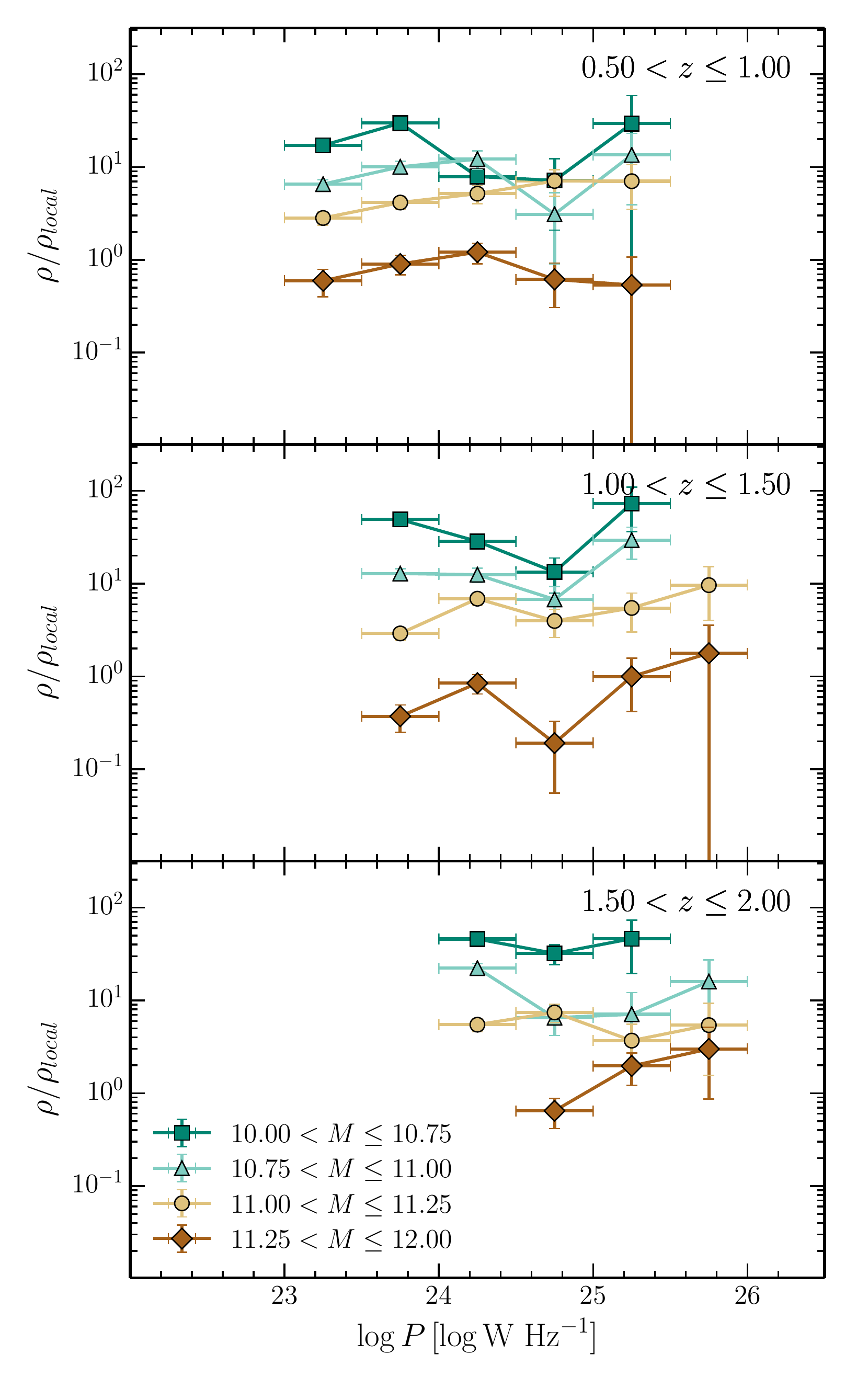}\\
 \caption{The relative comoving space density of radio-loud sources with respect to the local comoving space density. This is the RLF in redshift bins $0.5 < z \leq 1.0$ (top panel), $1.0 < z \leq 1.5$ (middle panel) and $1.5 < z \leq 2.0$ (bottom panel) by the local, $ 0.01 < z \leq 0.3$, RLF  for the four host stellar mass bins (plotted in colour).}
 \label{fig:rho_rholocal}
\end{figure}

\begin{figure}
 \centering
 \includegraphics[width=0.49\textwidth]{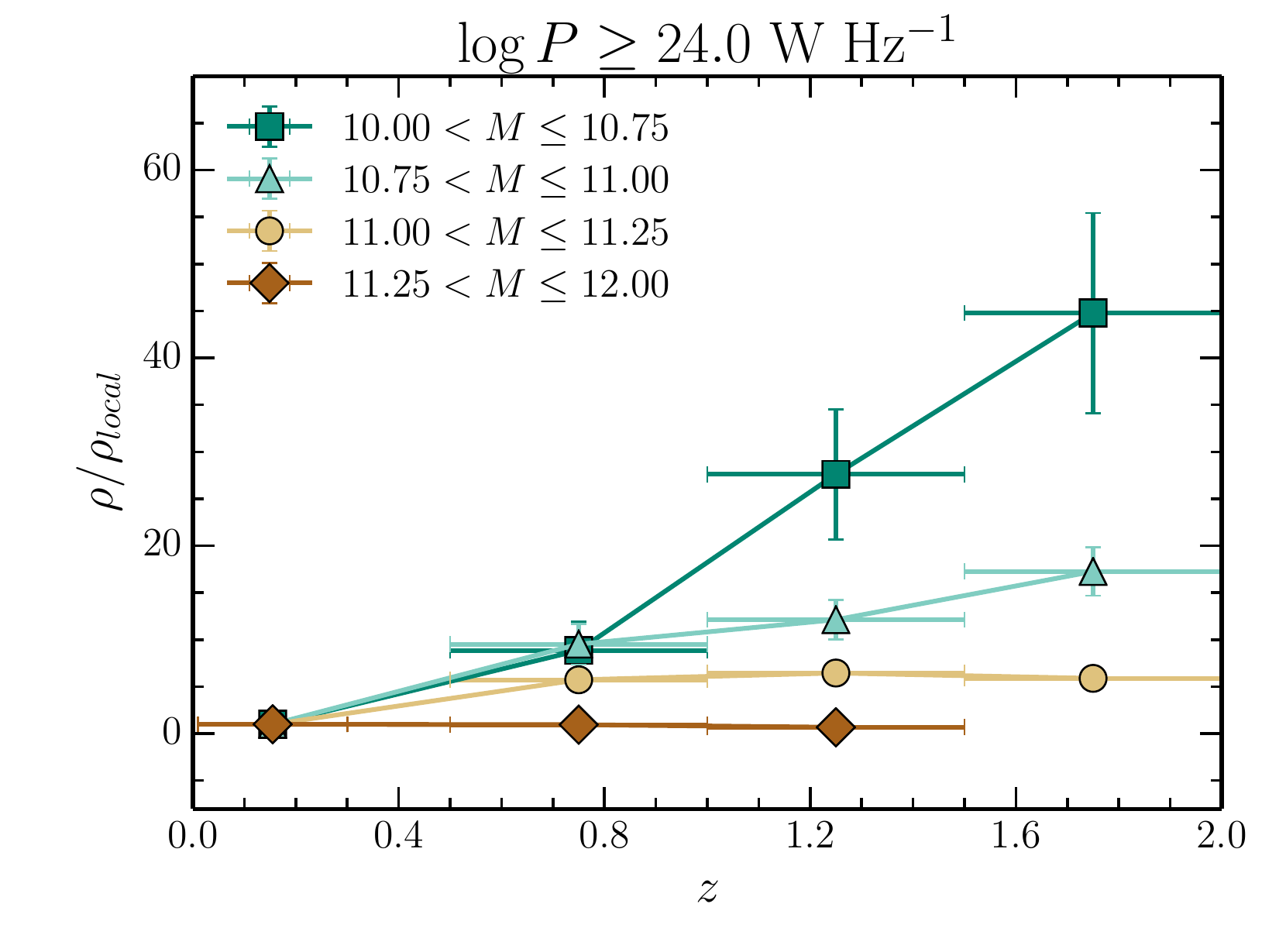}
 \caption{Relative space density of radio-loud sources with respect to the local density as a function of redshift for all sources with radio powers greater than the cut-off $\Pp > 10^{24}$\,\WHz. }
 \label{fig:rho_rholocal_vs_z}
\end{figure}

Now, to quantify the disproportional increase in radio activity among galaxies of different mass at higher redshifts, we investigate the relative comoving space density of radio-loud sources with respect to the local comoving space density. We do this by dividing the stellar mass dependent RLFs in redshift bins $0.5 < z \leq 1.0$, $1.0 < z \leq 1.5$ and $1.5 < z \leq 2.0$ by the local RLF ($ 0.01 < z \leq 0.3$). These relative RLFs are plotted in Fig.~\ref{fig:rho_rholocal}, from which it is clear that there is a difference between the relative comoving space density of low mass galaxies hosting radio sources and that of high mass hosts.  We derive the relative comoving space density of radio-loud sources with respect to the local comoving space density for all sources with radio powers greater than a  cut-off luminosity of  $\Pp > 10^{24}$\,\WHz. This is plotted as a function of redshift in Fig.~\ref{fig:rho_rholocal_vs_z} and the values listed in Table~\ref{tab:space_density_inc}.

{At $z\sim1.5-2$ the space density of the least massive galaxies hosting Radio-Loud AGN above  $\Pp > 10^{24}$\,\WHz is \emph{$45 \pm 11$} times greater than the  local space density. Even the sources with stellar masses in the range $ 10.75 <  \log (M_*/M_{\odot}) \leq 11.0 $ are $17.3 \pm 2.6$ times more prevalent. We note that there is a slight decrement in the space density of the most massive galaxies hosting Radio-Loud AGN at both radio power cuts going out to $z<1.5$. If we consider the slightly less powerful sources,  $\Pp > 10^{23.5}$\,\WHz, and only go out to redshift $z<1.5$ (not plotted here), the effect is only really seen in the lowest mass bin where the increase in space density {increases more rapidly with redshift}: the space density of the least massive galaxies hosting Radio-Loud AGN is \emph{$40.8 \pm 6.2$} times greater than locally compared to the $27.6 \pm 6.9$-fold increase for  $\Pp > 10^{24}$\,\WHz sources in the $1.0 < z \leq 1.5$ bin. }

\begin{table}
 \begin{center}
 \caption{{Relative space density of radio-loud sources with respect to the local density as a function of redshift for all sources with radio powers greater than the  cut-off  $\Pp > 10^{24}$\,\WHz.}}
 \label{tab:space_density_inc}
\begin{tabular}[c]{lccc}
\hline
\multicolumn{1}{c}{Mass bin} & \multicolumn{3}{c}{Redshift bin} \\
 \multicolumn{1}{c}{$\log (M_*/M_{\odot})$} & $0.5 - 1.0$ & $1.0 - 1.5$ & $1.5 - 2.0$ \\
\hline
 $ 10.00 - 10.75 $ & $8.9 \pm 3.1$ &       $27.6 \pm 6.9$ &      $45 \pm 11$  \\
 $ 10.75 - 11.00 $ & $9.5 \pm 2.2$ &       $12.1 \pm 2.1$ &      $17.3 \pm 2.6$  \\
 $ 11.00 - 11.25 $ &  $5.7 \pm 1.0$ &       $6.45 \pm 0.86$ &       $5.89 \pm 0.77$  \\
 $ 11.25 - 12.00 $ & $0.95 \pm 0.21$ &       $0.68 \pm 0.14$ &        \\ 
\hline
 \end{tabular}
 \end{center}
\end{table}

\section{The Radio-Loud Fraction}

Another way of looking at the increase in prevalence of AGN in lower mass hosts is to consider the fraction of sources which are radio-loud as a function of the host stellar mass in our four redshift bins. The mass-dependence of the radio-loud fraction can be an indicator of the accretion mode of the radio-AGN largely because of the different dependence of the fuelling source (hot vs. cold gas) on stellar mass \citep{Best.et.al.2006}.  The radio-loud fraction can be easily calculated by dividing the  stellar mass function (SMF) for radio-loud sources, $\rho_{Rad}$ (for sources above a given radio power limit), by the SMF for all galaxies, $\rho_{Opt}$:
\[f_{\mbox{radio-loud}} =\rho_{Rad}/ \rho_{Opt}.\]

\subsection{The Stellar Mass Function}

In order to calculate the radio-loud fraction we first derived the SMFs for all galaxies and for the radio source hosts  by using the $1/V_{max}$ estimator \citep{Schmidt.1968} as previously described for the LFs, which corrects for the fact that the samples are magnitude limited.  In order to construct the SMFs, the redshift-dependent limiting $M_{*}$ above which the magnitude-limited sample is complete needs to be known. For the COSMOS/UltraVISTA sample we use the empirical 95~per~cent completeness limit calculated by \cite{Muzzin.et.al.2013}. 

\begin{figure}
 \centering
 \includegraphics[width=0.49\textwidth]{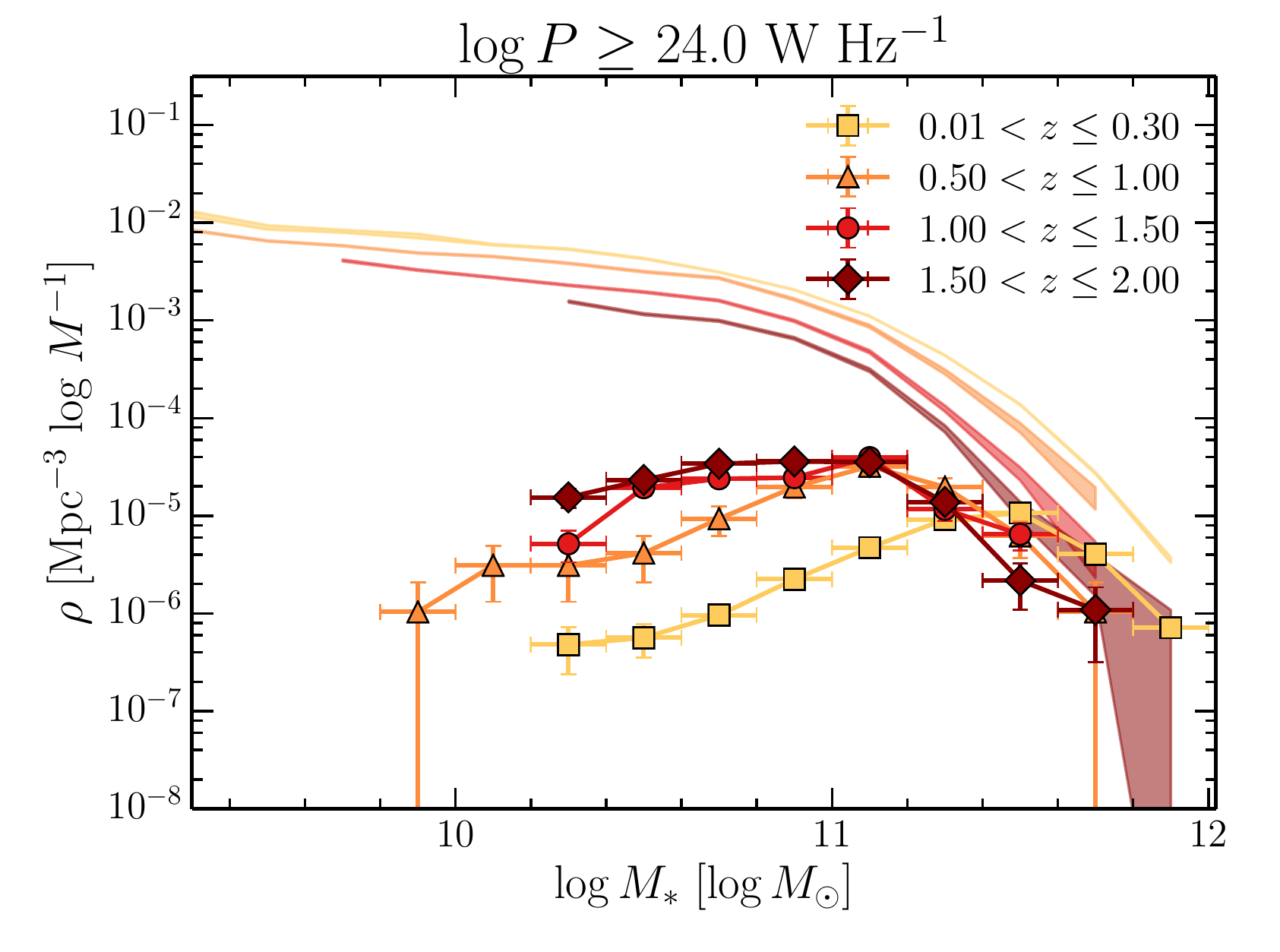}
 \caption{SMFs for all galaxies, $\rho_{Opt}$, (shaded lines) and radio-loud galaxies, $\rho_{Rad}$, (lines with points) for a radio-power cut-off $\Pp > 10^{24}$\,\WHz  in the four redshift bins.}
 \label{fig:smfs}
\end{figure}

The SMFs for radio-loud galaxies and all galaxies are plotted in Fig.~\ref{fig:smfs} for a radio-power  cut-off of  $\Pp > 10^{24}$\,\WHz. We choose this limit because the highest redshift bin is only able to probe radio powers greater than this (see also Fig.~\ref{fig:samplesboth}). The SMFs we derive for our optical galaxy samples (for both the SDSS and COSMOS samples) are consistent with those of \cite{Muzzin.et.al.2013b} who use a more sophisticated maximum-likelihood analysis to derive the SMFs in several redshift bins. As expected \citep{Best.et.al.2005b}, $\rho_{Rad}$ differs significantly from $\rho_{Opt}$ -- the hosts of radio sources are biased towards more massive systems. Interestingly, while the comoving number density of all galaxies decreases with redshift, that of the radio source hosts with $ \log (M_*/M_{\odot}) \la 11.0 $ increases. This is consistent with the results of \cite{Tasse.et.al.2008} and shows that the radio-loud galaxy population evolves differently from the population of galaxies as a whole.

\subsection{The Radio-Loud Fraction}

\begin{figure}
 \centering
 \includegraphics[width=0.49\textwidth]{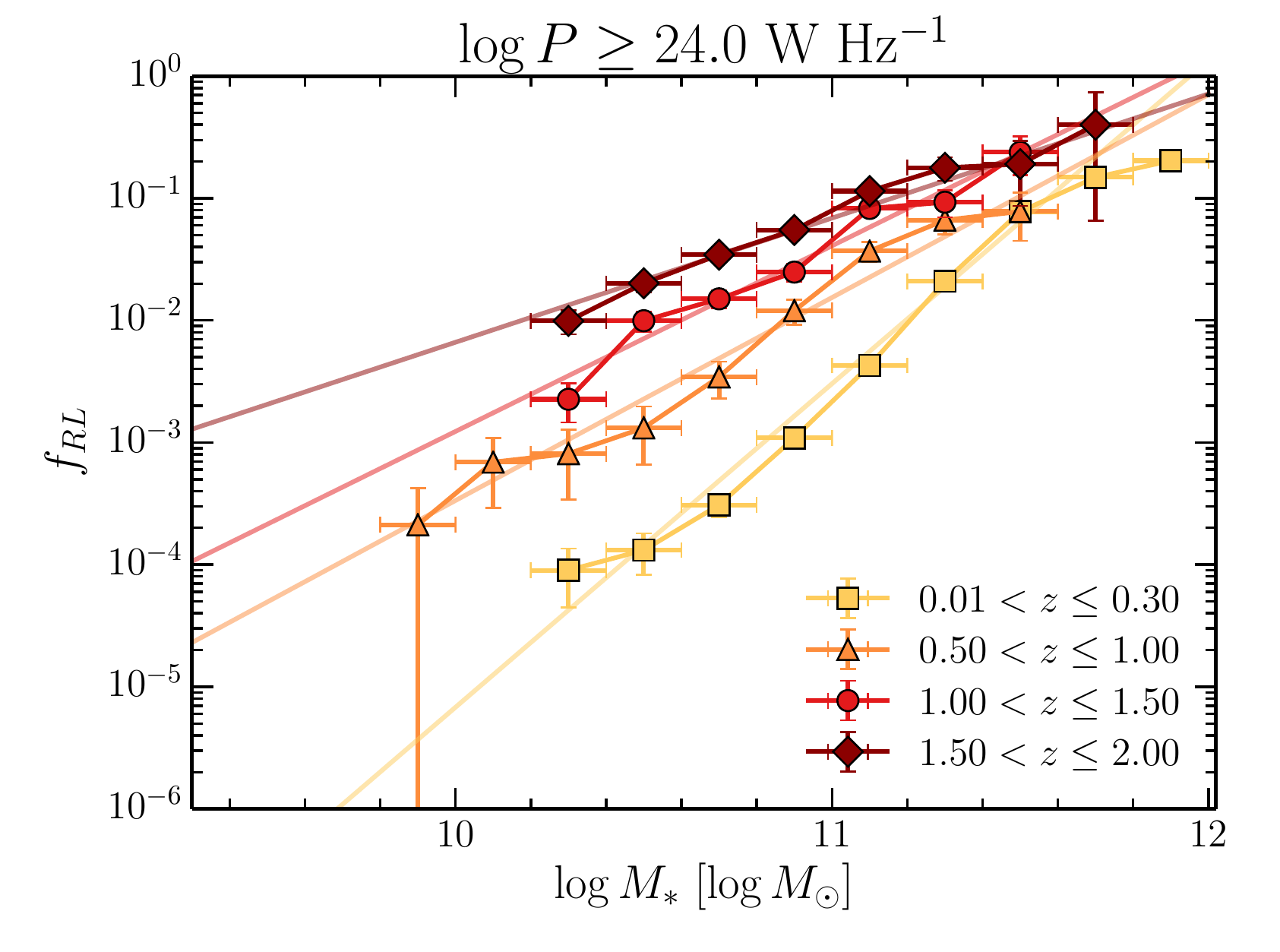}
 \caption{The fraction of galaxies hosting a radio source (radio-loud fraction) for a radio-power cut-off of  $\Pp > 10^{24}$\,\WHz  in the four redshift bins. The coloured lines show a linear fit over the stellar mass range $10 < \log (M_*/M_{\odot}) < 11.5 $. The slopes of these fits are $2.7 \pm 0.2$, $1.7 \pm 0.2$, $1.5 \pm 0.1$ and $1.0 \pm 0.1$ from the lowest to highest redshift bins.}
 \label{fig:rl_frac}
\end{figure}

\begin{figure}
 \centering
 \includegraphics[width=0.49\textwidth]{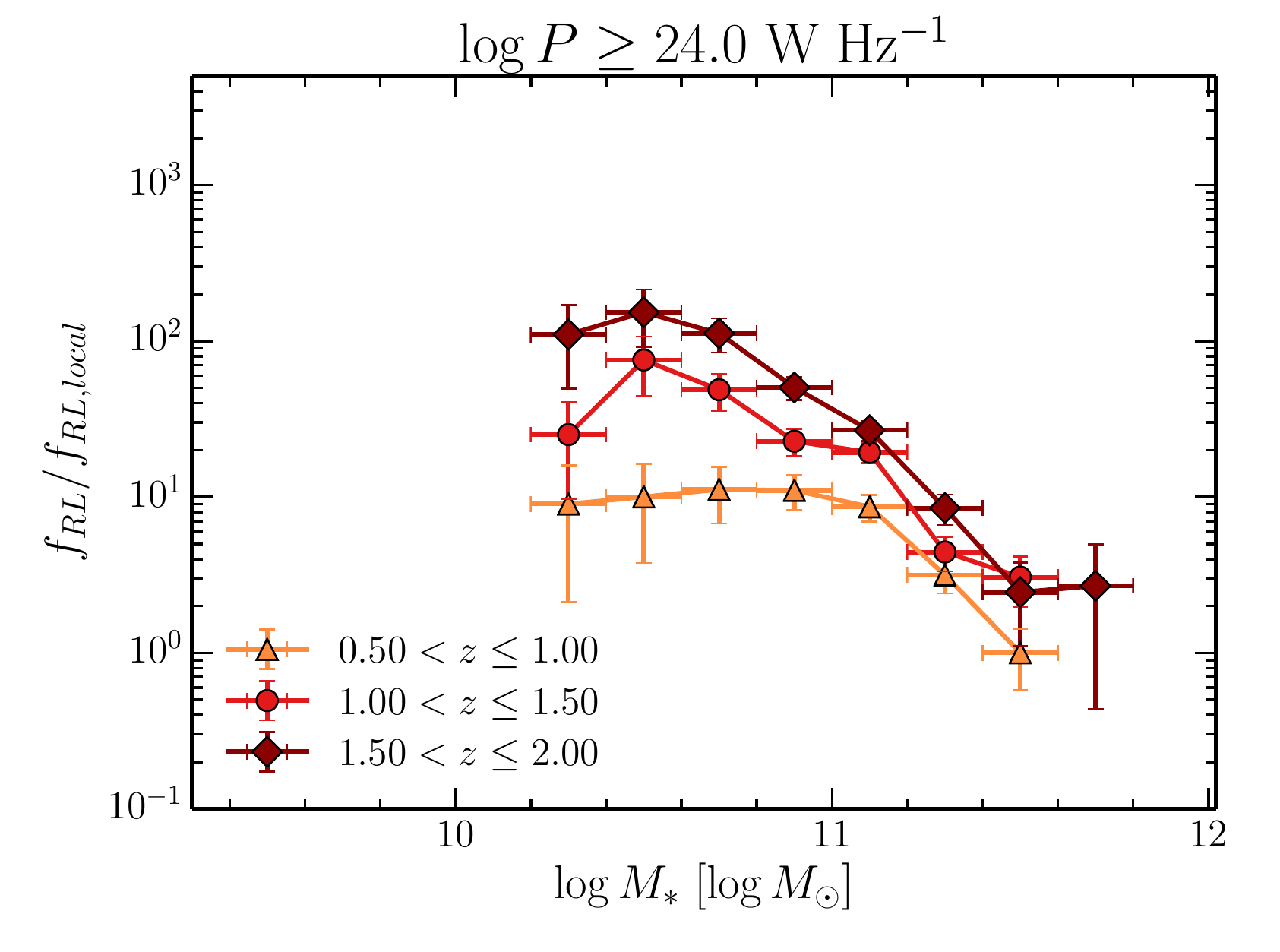}
 \caption{The fraction of galaxies hosting a radio source (radio-loud fraction) relative to the local fraction for a radio-power cut-off of $\Pp > 10^{24}$\,\WHz  in the three higher redshift bins.}
 \label{fig:rl_frac_local}
\end{figure}

Having computed the relevant SMFs we can calculate the radio-loud fraction as described above. Figure~\ref{fig:rl_frac} shows the radio-loud fraction  for a radio-power cut-off of $\Pp > 10^{24}$\,\WHz. The radio-loud fraction clearly increases with redshift. Moreover, the slope of the mass dependence becomes shallower, showing that the fraction of lower mass galaxies hosting radio sources increases more with redshift than the fraction for high mass galaxies.  The slopes are  shallower at higher redshifts, for $\Pp > 10^{24}$\,\WHz going from $f_{RL} \propto M_*^{2.7 \pm 0.2}$ in the local sample, flattening to  $f_{RL} \propto M_*^{1.7 \pm 0.1}$, $f_{RL} \propto M_*^{1.5 \pm 0.1}$ and $f_{RL} \propto M_*^{1.0 \pm 0.1}$  in the higher redshift bins.  This is highlighted in the plot of the radio-loud fraction relative to the local redshift bin (fig.~\ref{fig:rl_frac_local}), where the relative radio-loud fraction is up to two orders of magnitude greater at the low mass end.  We note that the flattening is quicker with redshift when the lower power $\Pp > 10^{23.5}$\,\WHz sources are considered (in the first three redshift bins only, not plotted here): $f_{RL} \propto M_*^{2.7 \pm 0.1}$, $f_{RL} \propto M_*^{1.3 \pm 0.1}$ and $f_{RL} \propto M_*^{1.1 \pm 0.1}$.

\section{Interpretation}
\label{sect:interpret}



In this section we aim to interpret and explain our results within the context of the HERG-LERG population dichotomy and their differential cosmic evolution and mass dependence.


{In the local Universe, the density of high mass Radio-Loud AGN is an order of magnitude higher than that of low mass Radio-Loud AGN at all radio powers (cf. top panel Fig.~\ref{fig:RLF_in_massbins}). In the more distant Universe, up to $z<2$ we see a sharp increase in the number density of Radio-Loud AGN hosted by lower mass galaxies, while the number density of high mass Radio-Loud AGN remains constant.}

{This large increase in the prevalence of radio activity among galaxies of lower mass at higher redshifts (cf. Fig.~\ref{fig:rho_rholocal_vs_z}) shows that it is the lower mass galaxies which are the cause of the upturn in the observed RLFs \citep[e.g.][]{Dunlop.Peacock.1990,Rigby.et.al.2011}. Moreover we suggest that this upturn is likely due to an increasing population of cold mode accretors at earlier epochs.  From \cite{Best.et.al.2005a} we know that locally, {despite the wide distributions in host stellar mass of both HERGs and LERGs, LERGs have a strong preference to be hosted by galaxies with higher masses, while HERGs are hosted by galaxies with a lower median stellar mass but with a broader distribution. Assuming this still holds at higher redshifts, it implies that this strongly evolving population of  lower mass Radio-Loud AGN are HERGs}. We also expect the highest mass, most powerful sources to peak in space density at higher redshifts  \citep[$z \sim 2-3$][]{Rigby.et.al.2011} in line with the cosmic downsizing picture where the most massive black holes have formed by $z\sim4$. Indeed, results from many of the earlier radio surveys show that the most powerful ($\Pp \gtrsim 10^{26}$\ \WHz) radio galaxies at $z \gtrsim 1$ \citep[e.g.][]{Eales.et.al.1997,Jarvis.et.al.2001b,Seymour.et.al.2007,Fernandes.et.al.2015}  are predominantly HERGs hosted by the most massive galaxies.}

Indications from studies out to $z \lesssim 1$ show that the HERGs are indeed evolving more strongly with redshift than the LERG population \citep{Best.et.al.2014} such that within the redshift ranges of this study the radio-AGN population should be dominated by HERGs, opposite to that within the local universe. The mass dependence that we observe in this study supports this idea. Moreover, the evolution in the optical quasar luminosity function  
\citep[i.e. that corresponding to radio-quiet cold mode accretion;][]{Hasinger.et.al.2005,Hopkins.et.al.2007,Croom.et.al.2009} is comparable to the kind of increase  we observe for the low mass galaxies. 

Furthermore, the slope of the radio-loud fraction (cf. Fig.~\ref{fig:rl_frac}) we observe at the higher redshifts, $f_{RL} \propto M_*^{\sim1.3}$, is more consistent with the slope of the radio-loud fraction found in the local universe for HERGs only \citep{Janssen.et.al.2012}. On the other hand, in the local sample, our derived slope of  $f_{RL} \propto M_*^{\sim2.5}$ matches that which \cite{Janssen.et.al.2012} found for only LERGs and matches the theoretical value for the accretion of hot gas from a halo \citep{Best.et.al.2006}. In this case we know that the dominant population of all local radio sources is that of the LERGs.  This lends support to the idea that there is an increase in the prevalence of HERG activity or cold mode accretion and that this mode is becoming the dominant population out to redshifts of $0.5 < z \leq 2$.

Lastly, as \cite{Heckman.Best.2014} state, the crucial requirement for cold mode accretion is the abundant central supply of cold dense gas. And as reflected in the increase in cosmic star formation rate density which has increased tenfold out to $z \sim 2$ and threefold out to $z \sim 0.5$ \citep[e.g.][and references therein]{Sobral.et.al.2013, Madau.Dickinson.2014}, there is significantly more cold gas fuelling star formation at these epochs. 

A closer inspection of Fig.~\ref{fig:rho_rholocal} reveals some interesting features, most particularly in highest mass bin, $11.25 < \log (M_*/M_{\odot}) < 12$, in the highest redshift interval, $1 < z \leq 1.5$. We suggest that while the radio-AGN sample as a whole at these redshifts is becoming dominated by cold mode accretors, the highest mass, intermediate power ($10^{24} < \Pp <10^{25}$\,\WHz) sources  could still be  hot mode/LERG sources because of their high mass. We observe a slight decrease in the number density of these specific sources consistent with the modelling results of \citeauthor{Rigby.et.al.2015} where the $\sim10^{24}$\,\WHz population peaks at $z\sim1$ and is likely dominated by LERGs. This could be significant in the context of the models of LERG evolution presented by \cite{Best.et.al.2014}. If these are indeed LERGs the decrease in number density does not match the order of magnitude decrease predicted by  the preferred model of \cite{Best.et.al.2014} extrapolated out to $z \sim 1.5$ from fits to data out to $z \sim 0.7$. This model includes a time delay of $\sim 1.5$\,Gyr between the formation of massive quiescent galaxies and when they are able to produce hot mode AGN, and at redshifts above $\sim 1$ the available population of host galaxies is declining so rapidly that such a delay in the onset of hot mode AGN activity would imply a drastic fall in the number densities of these sources above $z \sim 1$.

\section{Summary and Conclusions}
\label{sect:conclude}

In this paper we have used  the SDSS value-added spectroscopic sample of radio-loud galaxies \citep{Best.Heckman.2012} and the  VLA-COSMOS radio sample \citep{Schinnerer.et.al.2004, Schinnerer.et.al.2007} matched to a $K_s$-selected catalogue of the COSMOS/UltraVISTA field \citep{Muzzin.et.al.2013} to compile two samples of Radio-Loud AGN going out to $z=2$. The samples are of sufficiently high radio power that they are  dominated by RL AGN. Using these samples we have
constructed radio luminosity functions for four host stellar mass bins between $ \log (M_*/M_{\odot}) = 10.0$ and $\log (M_*/M_{\odot}) =12.0 $, in four redshift bins between $z=0.01$ and $z=2$. We have also investigated the radio-loud fraction as a function of stellar mass in these redshift bins. Together, we found the following:
\begin{enumerate}
 \item  Radio activity among galaxies of different mass increases differently towards higher redshifts. By considering the relative comoving space density of radio-loud sources with respect to the local comoving space density, we showed that at $1.5 < z < 2$ the space density of galaxies with stellar masses in the range $ 10.00 <  \log (M_*/M_{\odot}) \leq 10.75 $ and $ 10.75 <  \log (M_*/M_{\odot}) \leq 11.0 $   hosting Radio-Loud AGN with {$\Pp >10^{24}$\,\WHz is respectively \emph{$45 \pm 11$} and \emph{$17.3 \pm 2.6$}} times greater relative to the local space density while that of higher mass galaxies hosting Radio-Loud AGN remains the same.
 \item  The fraction of galaxies which host Radio-Loud AGN with $\Pp > 10^{24}$\,\WHz as a function of stellar mass  shows a clear increase with redshift and  a flattening with mass with the mass dependence evolving from $f_{RL} \propto M_*^{2.7}$ in the local sample to  $f_{RL} \propto M_*^{1.0}$ at $1.5 < z < 2$.
\end{enumerate}

We have argued that this increase in the prevalence of radio activity among galaxies of lower mass at higher redshifts is largely due to a rising contribution of AGN accreting in the radiative mode (HERGs). With this data we cannot yet conclusively show the evolution of the different accretion modes as a function of their host stellar mass because we lack the information on the excitation state of these sources at higher redshifts. However, future work combining this and other radio-optical samples will allow more detailed studies of optical hosts of the high-redshift ($1 \lesssim z \lesssim 2$) population of radio-AGN.

\section*{Acknowledgements}

The authors thank Philip Best, George Miley and Emma Rigby for useful discussions which contributed to the interpretation in this paper. {The authors thank the anonymous referee for useful comments which have improved this manuscript}.

This study uses a K$_{s}$-selected catalogue of the COSMOS/UltraVISTA field from \cite{Muzzin.et.al.2013}.  The catalogue contains PSF-matched photometry in $30$ photometric bands covering the wavelength range $0.15\micron$ $\rightarrow$ $24\micron$ and includes the available $GALEX$ \citep{Martin.et.al.2005}, CFHT/Subaru \citep{Capak.et.al.2007}, UltraVISTA \citep{McCracken.et.al.2012}, S-COSMOS \citep{Sanders.et.al.2007}, and zCOSMOS \citep{Lilly.et.al.2009} datasets.  This work uses the catalogue compiled by \cite{Best.Heckman.2012}, which combines data from the VASC \citep{Brinchmann.et.al.2004} of the SDSS DR7 \citep{Abazajian.et.al.2009} with NVSS \citep{Condon.et.al.1998} and FIRST \citep{Becker.et.al.1995}.

\bibliographystyle{mn2e}
\bibliography{lfbibfile}

\bsp

\label{lastpage}

\end{document}